\documentclass[a4paper]{jpconf}
\usepackage{graphicx, amsfonts}
\begin{document}
\title{A quantum field theory of simplicial geometry and the emergence of spacetime}

\author{Daniele Oriti}

\address{Institute for Theoretical Physics and Spinoza Institute, Utrecht University \\ Minnaert building, Leuvenlaan 4, Utrecht, The Netherlands}

\ead{d.oriti@phys.uu.nl}

\begin{abstract}
We present the case for a fundamentally discrete quantum spacetime
and for Group Field Theories as a candidate consistent description
of it, briefly reviewing the key properties of the GFT formalism.
We then argue that the outstanding problem of the emergence of a
continuum spacetime and of General Relativity from fundamentally
discrete quantum structures should be tackled from a condensed
matter perspective and using purely QFT methods, adapted to the
GFT context. We outline the picture of continuum spacetime as a
condensed phase of a GFT and a research programme aimed at
realizing this picture in concrete terms.
\end{abstract}

\section{Introduction: quantum gravity, discreteness and the problem of the continuum}
What is the problem of quantum gravity? The most straightforward
and naive answer is that it is to construct a quantum field theory
of the gravitational field, obtained by quantizing somehow the
corresponding classical field theory. However, since the classical
theory is General Relativity, the above answer is unsatisfactory.
The key insight of GR is that the gravitational field is to be
understood as spacetime geometry, and to quantize it means
understanding what it means to have a \lq quantum spacetime
geometry\rq. Even more, given the strict relation between
spacetime geometry and topology, it has long been suggested that a
quantization of geometry should involve a dynamical spacetime
topology as well. Therefore, a theory of quantum gravity aims to
be not just a theory of a certain physical interaction, but the
codification of a new understanding of {\bf what spacetime is} at
the most fundamental (quantum and microscopic) level \cite{libro}.

\subsection{Spacetime discreteness, the \lq atoms of space\rq and the problem of the continuum}
One fascinating possibility is that, at high energies and small
distances, spacetime loses its continuum appearance and is instead
best described in terms of discrete structures. This possibility
is a basic assumption in some approaches to quantum gravity, e.g.
causal set theory \cite{causalset}, as a convenient computational
tool in others, e.g. simplicial quantum gravity \cite{renate,
ruth}, and it is the natural (preliminary) outcome of standard
quantization techniques applied to GR, e.g. in Loop Quantum
Gravity \cite{lqg}, being also suggested by several results in
string theory.

We refer to the literature for all the arguments leading to the
hypothesis of spacetime discreteness (e.g.\cite{rafael}). Among
them, we can mention the many results establishing the laws of
black hole thermodynamics \cite{wald}. A thermodynamics is usually
the macroscopic encoding of an underlying microscopic statistical
mechanics for the system, i.e. black holes, governing the dynamics
of its microscopic degrees of freedom. The finiteness of black
hole entropy further suggests the discreteness of at least the
degrees of freedom constituting the horizon of the black hole (to
which the entropy is associated). But a black hole is just a
particular region of spacetime, and it is therefore the
statistical mechanics of spacetime itself that we are forced to
unravel, and express in terms of some unknown microscopic discrete
constituents. Another body of arguments suggests a special role
for the Planck length $l_p = \sqrt{h G/c^3}$ in any future theory
of quantum gravity \cite{planck}. In absence of such a complete
theory, of course, such arguments remain just motivations, but
certainly the appearance of the Planck length as either a minimal
length or as a minimal resolution for geometric measurements would
prove the usual description of spacetime as a continuum inadequate
at the more fundamental quantum level.

Now, to what extent do existing approaches support a discrete
picture of spacetime? what description do they give of the
hypothetical building blocks, the \lq atoms of space\rq?

Kinematical states of the gravitational field in Loop Quantum
Gravity \cite{lqg} are given by {\bf spin networks}, graphs whose
links are labeled by representations of the symmetry group
chosen, and whose vertices are labeled by intertwiners of the
same group. Thus spacetime as a continuum disappears at the
quantum level and is replaced by purely combinatorial and
algebraic degrees of freedom. In turn, any spin networks can be
thought of as built from the composition of their individual
vertices (better, of open spin networks with a single vertex each)
at the combinatorial level, and of the corresponding intertwiners
at the algebraic level. So one could say that, as far as LQG is
concerned, the \lq atoms of space\rq are group intertwiners or
spin network vertices. Let us consider now simplicial quantum
gravity approaches, by which we mean both quantum Regge calculus
\cite{ruth} and dynamical triangulations \cite{renate}. While the
way these two approaches encode the geometric degrees of freedom
of spacetime is different, in both cases spacetime is represented
by a simplicial complex, at the quantum level, and a continuum
D-dimensional spacetime emerges from the dynamics of a (infinite)
collection of (D-1)-simplices, which can be interpreted as the \lq
atoms of space\rq. In the most studied spin foam models
\cite{alex}, strictly related to GFTs, the atoms of space are spin
networks, as in LQG, but the graph on which these spin networks
are based is dual to a simplicial complex; thus these atoms can be
represented, in spin foam models, and GFTs, both as group
intertwiners and as fundamental simplices.

In all the above mentioned approaches, just as in {\it any}
current discrete approach to quantum gravity, the outstanding
problem is that of the continuum: given a fundamental description
of spacetime as made out of discrete quantum building blocks, and
a well-defined quantum dynamics for them, can we show that this
dynamics leads to a continuum description of spacetime, and is
effectively described by General Relativity (or some modification
of it) in some (low energy/large scale) approximation? How does
the continuum {\it emerge} from a fundamentally discrete picture?
How does the dynamics of General Relativity emerge from a
microscopic dynamical theory that does not refer to the continuum
at all, and likely neither does refer to geometry in the first
place? This entails a kinematical question, that of the
approximation of the fundamentally discrete structures with
continuum ones, but also as a dynamical process, in which the
evolution of the fundamentally discrete system leads, under
certain conditions and to some approximation, to a configuration
where the continuum description is possible. The same problem is
faced, although of course it manifests itself and it is tackled in
a different way, in all discrete approaches. Each of them has
developed techniques and ideas for tackling this problem: from
weave states and \lq statistical geometry\rq techniques in LQG
\cite{lqg, bombelli}, to refinement and lattice renormalization
group techniques in simplicial quantum gravity \cite{renate}, to
the large body of methods of causal set theory \cite{causalset}.
In spite of this intense activity, and also of the many promising
results (e.g. \cite{renate}), it is fair to say that the problem
of the continuum remains wide open.

\subsection{An emergent spacetime? The condensed matter analogy}
Concerning the problem of the continuum in quantum gravity, we
feel that the results obtained in condensed matter analogs of
gravitational phenomena \cite{volovik, hu} are very inspiring, and
suggest a perspective that is particularly suited for group field
theories. Let us summarize some of these results. We will be of
course very sketchy, and also consider only one example of a
condensed matter system that is of interest for gravitational
physics: $^3 He$. We refer to the vast literature on the subject,
and in particular to \cite{volovik}, for a more extensive,
detailed and competent review.

$^3 He$ is made of fermionic atoms. The \lq fundamental\rq  theory
at work for this system, its own \lq theory of everything\rq
\cite{laughlin}, is therefore an interacting relativistic field
theory of fermions. Such description is however useless for all
practical purposes, and the system is perfectly described in the
non-relativistic approximation. At temperatures well above the
superfluid transition temperature $T_c$, the system is a gas; as
the temperature drops, but remaining above $T_c$, the atoms
condense to a liquid phase. It can be described by the action:

\begin{equation}
S(\psi)\,=\,\int dt d^3x \left[ \psi^{\dagger}_\alpha \left(
i\,\partial_t\, -\,\frac{p^2}{2m}\,+\,\mu
\right)\psi_{\alpha}\right] \,+\, S_{\textit{int}} \label{eq:He}
\end{equation}

where $\psi$ is the fundamental field operator for the $^3 He$
atoms with mass $m$, $\mu$ is the chemical potential and
$S_{\textit{int}}$ is the quartic term describing the pair
interaction of the atoms. It is a system with a Fermi surface,
described by a very large number of degrees of freedom. As the
temperature drops further, however, the number of degrees of
freedom is drastically reduced and the system can be described
solely in terms of non-interacting fermionic quasi-particles above
the Fermi surface. This effective theory is valid for temperatures
well below the Fermi temperature $\Theta = \frac{\hbar^2}{m a^2}$
($a$ is the average inter-particle distance in the liquid), which
plays the role of the \lq Planck energy\rq, and can be dealt with
in detail using the BCS methods \cite{volovik}. At this level we
can use an approximate hydrodynamic description of the liquid,
expressing its dynamics in terms of collective liquid variables,
e.g. the velocity field $v_s(x)$, and quasi-particles degrees of
freedom, i.e. collective propagating excitations of the liquid,
while the liquid itself is well-approximated as a continuum. The
BCS treatment shows that the system belongs to the universality
class of Fermi systems with a Fermi point (in the so-called \lq
A-phase\rq), and the quasi-particles possess, in the vicinity of
the Fermi point, the effective \lq relativistic\rq dispersion
relation:

$$
g^{\mu\nu}(p_\mu - e A_\mu)(p_\nu - e A_\nu)\,=\,0
$$
in which one has introduced an effective metric and an effective
electromagnetic field, as seen by the propagating quasi-particles.
We are concerned here only with the effective metric, given by the
collective variables of the liquid-continuum: the superfluid
velocity $v_s^i$ and the $l$-field, which measures the vorticity
of $^3 He-A$, or, in geometric terms, the anisotropy of space, as:
$$
g_{ij}=\frac{1}{c_{||}^2} l^i l^j +
\frac{1}{c_{\perp}^2}(\delta^{ij} - l^i
l^j)\;\;\;\;g_{00}=-(1-g_{ij} v^{i}_s
v^{j}_s)\;\;\;\;g_{0i}=-g_{ij} v_s^{j} \;\;\;\;
\sqrt{-g}\,=\,\frac{1}{c_{||}c_{\perp}^2},
$$
where the parameters $c_{||}$ and $c_{\perp}$ arise from the
microscopic physics \cite{volovik}, and play the role of the
velocity of light in the direction parallel and orthogonal to $l$
respectively.
 The
effective action for the collective variables is composed of
various terms, some of them depending on the superfluid velocity
field $v_s$, others depending on the $l$-field instead. In
general, the terms depending on $v_s$ do not have an
interpretation in terms of a gravitational theory and do not
relate easily to General Relativity; at the same time they tend to
dominate over those, depending on the $l$-field, that do have such
interpretation. However, these terms are suppressed in the limit
of heavy atoms, since $v_s\sim 1/m$, i.e. the limit of inert
vacuum. In this limit, together with other terms which have an
electromagnetic interpretation, we are left with an effective
action proportional to: $\int d^3x \,(l \cdot (\nabla \times l))^2
$, which in turn is nothing but $\frac{1}{16\pi G}\int \sqrt{-g}\,
R(g)$, i.e. the Einstein action, for the effective metric
$g_{\mu\nu}(l)$ in which the $v_s$-dependent terms have been
dropped, and the effective Newton constant can be expressed in
terms of the microscopic parameters of the superfluid.

Let us summarize. We have a discrete system whose building blocks
are fermionic atoms and a microscopic field theory describing its
dynamics. At low temperature, the system undergoes a phase
transition and condenses to a liquid phase. In this phase most of
the microscopic details are irrelevant (not all of them, of
course; e.g. if the system was a made of atoms living in 6 space
dimensions, say, the collective velocity of the fluid would be
6-dimensional, or in $^4 He$ vorticity is absent and a different
class of metrics emerge, etc), and one can adopt an effective
hydrodynamic description in terms of a continuum, and of
collective variables describing the fluid, together with low
energy excitations (quasi-particles) propagating in it. The
microscopic theory provides the values of the effective \lq
fundamental constants\rq of the macroscopic theory. The collective
variables of the liquid behave like an effective metric field (and
other interaction fields as well), and their dynamics admits a
description, at least in some limit, in terms of the Einstein
action. The theory is thus approximately generally covariant
(internal symmetries emerge as well). This behaviour is only
approximate (preferred directions can be identified) and only
valid at very low temperature; moreover, the class of metrics that
emerge is large but does not include all the configuration space
of General Relativity. These limitations are a consequence,
ultimately, of the non-relativistic nature of the fundamental
atomic system considered, and of other details of it. These
limitations notwithstanding, we have a concrete example of how: 1)
the continuum can be understood as a convenient, if not necessary,
approximation of a fundamentally discrete system; 2) spacetime and
of geometry can {\bf emerge} from a theory that is not about
spacetime nor geometry nor gravity; 3) General Relativity itself
can {\bf emerge} as an effective description of the
(hydro-)dynamics of the collective continuum variables of a
microscopically discrete system.

How these results should change our views about spacetime and
gravity is of course a matter of debate \cite{volovik, hu}. It
seems to us that these results fit nicely with the point of view
outlined in the beginning, which sees a continuum spacetime as an
approximation of some yet to be discovered \lq atoms of space\rq,
described by a theory that is not expressed in terms of a
pre-existing spacetime to start with, and from which General
Relativity emerge in a low energy or macroscopic limit. The
question then is not of principles, but very practical: can we
make a more direct use of the insights provided by condensed
matter systems? can we identify a context in which these ideas and
techniques can be directly applied? We will argue below that Group
Field Theories can represent such a context, and outline a
research programme aimed at implementing the tools and ideas of
condensed matter gravity analogs in quantum gravity, solving in
the process the outstanding difficulties of existing discrete
quantum gravity approaches.

\section{Group Field Theories}
We now give a very brief sketch of Group Field Theories, referring
to the literature for more details \cite{iogft, laurentgft}. GFTs
can be seen, on the one hand, as a generalization of matrix models
for 2d quantum gravity to higher dimensions, and on the other hand
as a complete definition of spin foam models in that they provide
a natural prescription for the sum over spin foam 2-complexes that
is necessary to capture in full the dynamics of these models; as
such they share ideas and mathematical structures with both
simplicial approaches and Loop Quantum Gravity. GFTs can also be
understood as a local and discrete realization of the idea of a
3rd quantization of gravity, including a sum over topologies
alongside a covariant sum over geometries for given topology. For
GFTs aimed at describing D-dimensional quantum gravity, the field
is a $\mathbb{C}$-valued function of D group elements
$\phi(g_1,..,g_D)$, for a generic group $G$ being either the
$SO(D-1,1)$ Lorentz group (or $SO(D)$ for Riemannian gravity), or
some extension of it. It can be interpreted as a second quantized
(D-1)-simplex, and each argument corresponds to one of its
boundary (D-2)-faces. The closure of the D (D-2)-faces to form a
(D-1)-simplex is encoded in the invariance under diagonal action
of the group $G$: $\phi(g_1,...,g_D)=\phi(g_1g,...,g_Dg)$. The
mode expansion gives:
$$\phi(g_i)=\sum_{J_i,\Lambda,k_i}\phi^{J_i\Lambda}_{k_i}\prod_iD^{J_i}_{k_il_i}(g_i)C^{J_1..J_D\Lambda}_{l_1..l_D}, $$ with the $J$'s
labeling representations of $G$, the $k$'s vector indices in the
representation spaces, and the $C$'s being intertwiners of the
group $G$, labeled by an extra parameter $\Lambda$.  Group
variables represent configuration space, while the representation
parameters label the corresponding momentum space. Geometrically,
the group variables represent parallel transport of a connection
along elementary paths dual to the (D-2)-faces, while the
representations $J$ can be put in correspondence with the volumes
of the same (D-2)-faces, the details of this correspondence
depending on the specific model. A simplicial space built out of
$N$ such (D-1)-simplices is then described by the tensor product
of $N$ field operators, with suitable contractions implementing
the fact that some of the (D-2)-faces are identified. States of
the theory are then labeled, in momentum space, by {\it spin
networks} of the group $G$. Spacetime, represented by a
D-dimensional simplicial complex, emerges in the perturbative
expansion of the GFT partition function, as a particular
interaction process among (D-1)-simplices, i.e. as a GFT Feynman
diagram. The action is chosen, with this goal in mind, to be of
the form:
$$
S= \frac{1}{2}\int
  dg_id\tilde{g}_i\,
  \phi(g_i)\mathcal{K}(g_i\tilde{g}_i^{-1})\phi(\tilde{g}_i)
  +
  \frac{\lambda}{(D+1)!}\int dg_{ij}\,
  \phi(g_{1j})...\phi(g_{D+1 j})\,\mathcal{V}(g_{ij}g_{ji}^{-1}),
$$
where the choice of kinetic and interaction functions
$\mathcal{K}$ and $\mathcal{V}$ define the specific model. The
interaction term describes the interaction of D+1 (D-1)-simplices
to form a D-simplex by gluing them along their (D-2)-faces
(arguments of the fields). The nature of this interaction is
specified by the choice of function $\mathcal{V}$. The kinetic
term involves two fields each representing a given (D-1)-simplex
seen from one of the two D-simplices (interaction vertices)
sharing it, so that the choice of kinetic functions $\mathcal{K}$
specifies how the geometric degrees of freedom corresponding to
their D (D-2)-faces are propagated from one vertex of interaction
(fundamental spacetime event) to another. A GFT is an almost
ordinary field theory, with a fixed background metric structure
and the usual splitting between kinetic (quadratic) and
interaction (higher order) term in the action, allowing for a
straightforward perturbative expansion. However, the action is
also {\it non-local} in that the arguments of the D+1 fields in
the interaction term are not all simultaneously identified, but
only pairwise. Most of the work up to now has focused on the
perturbative aspects of quantum GFTs, using the expansion in
Feynman diagrams of the partition function:

$$ Z\,=\,\int
\mathcal{D}\phi\,e^{-S[\phi]}\,=\,\sum_{\Gamma}\,\frac{\lambda^N}{sym[\Gamma]}\,Z(\Gamma),
$$
where $N$ is the number of interaction vertices in the Feynman
graph $\Gamma$, $sym[\Gamma]$ is a symmetry factor for the graph
and $Z(\Gamma)$ the corresponding Feynman amplitude. Each edge of
the Feynman graph is made of $D$ strands, one for each argument of
the field, and each one is then re-routed at the interaction
vertex, following the pairing of field arguments in the vertex
operator. Each strand goes through several vertices, coming back
where it started, for closed Feynman graphs, and therefore
identifies a 2-cell. Each Feynman graph $\Gamma$ is then a
collection of 2-cells (faces), edges and vertices, i.e. a
2-complex, that, because of the chosen combinatorics for the
arguments of the field in the action, is topologically dual to a
D-dimensional simplicial complex. Clearly, the resulting
complexes/triangulations can have arbitrary topology, each
corresponding to a particular {\it scattering
  process} of the fundamental building blocks of space,
i.e. (D-1)-simplices. In momentum space, each Feynman graph is
given by a spin foam (a 2-complex with faces $f$ labeled by
representation variables), and each Feynman amplitude (a complex
function of the representation labels) by a spin foam model: $
Z(\Gamma)=\sum_{\{J_f\}} A(\{J_f\}).$ The representation variables
have a geometric interpretation (edge lengths, areas, etc) and so
each of these Feynman amplitudes defines a sum-over-histories for
discrete quantum gravity on the specific triangulation dual to the
Feynman graph. At the same time, this sum over geometries is
generated within a sum over simplicial topologies corresponding to
the perturbative sum over Feynman diagrams. The transition
amplitude between certain boundary data represented by two spin
networks, of arbitrary combinatorial complexity, can be expressed
as the expectation value of the field operators having the same
combinatorial structure of the two spin networks. Moreover, the
restriction of the GFT perturbative expansion to {\it tree level},
involving then only {\it classical GFT information} and generating
simplicial spacetimes of trivial topology only, for given boundary
spin networks, can be considered \cite{laurentgft} as the GFT
definition of the canonical inner product, implementing the action
of the Hamiltonian constraint operator on spin network states.

 Most of the model building has been based on the description
of classical gravity as a topological BF theory in 3d , or as a
constrained BF theory in higher dimensions,, and leads to very
simple GFT models, with kinetic and vertex terms given just by a
product of delta functions over the group $G$ or over suitable
homogeneous subspaces of it:
$$ \mathcal{K}(g_i,\tilde{g}_i) = \int_G dg \prod_i
\delta(g_i\tilde{g}_i^{-1}g),\;\;\;\;\;\;\mathcal{V}(g_{ij},g_{ji})
= \prod_i\int_G dg_i \prod_{i<j}\delta(g_i
g_{ij}g_{ji}^{-1}g_j^{-1}),
$$
where the integrals impose the $G$-invariance. These models have
thus a very simple structure, and can be motivated in more than
one way as candidate descriptions of quantum gravity. At the same
time, they are rather peculiar field theories, because, on top of
their non-locality, they lack the usual derivative operators in
the kinetic term.

Much more is known about the general structure of GFTs, and about
the various specific GFT models that have been constructed up to
now. For all this, we refer again to the literature.

GFTs, just as matrix models in the 2d case, provide thus a picture
of a discrete spacetime as {\bf emergent} as a Feynman graph from
a theory that is not about spacetime at all. While remarkable,
this is of course not the solution to the problem of the
continuum, as outlined above. For this, a different perspective is
needed, and we will outline it in the next section.

Before doing so, however, we want to mention another potentially
important feature of the GFT formalism. This is the possibility to
identify, within the GFT approach, many of the ingredients and
structures that are present in other discrete approaches to
quantum gravity. Boundary states are spin networks, as in LQG, but
at the same time they have a dual description as simplicial
spaces, as in simplicial quantum gravity. Their dynamics is
expressed as a covariant sum over geometries as in spin foam
models, or a discrete gravity path integral, as in quantum Regge
calculus, but involves also a sum over inequivalent
triangulations, as in the dynamical triangulations approach. The
histories summed over are GFT Feynman diagrams, which are directed
graphs or pre-orders from a mathematical point of view,
interpreted as collection of fundamental spacetime events linked
by fundamental causal relations, thus sharing some structures and
ideas with the causal set approach to quantum gravity. More links
can be found, as well as more work is needed to clarify or
strengthen such links, as discussed, for example in \cite{iogft,
generalised}, but we believe that GFT can be useful in providing
bridges between these various approaches and for understanding
them from a single, although un-conventional, perspective. A key
step toward this goal would be to construct a GFT that has
quantum amplitudes having the form of the exponential of a
classical action, say the Regge action for simplicial gravity.
Work on this is in progress, and a candidate model
\cite{generalised} is based on the following kinetic term: $$
K\,=\,\prod_{i=1}^{D}\int_G dg_i\,\int_\mathbb{R}ds_i \left\{
\phi^{*}(g_i,s_i)\left[\prod_i\left(
  -i\partial_{s_i}+\nabla_{i}\right)\right]\phi(g_i,s_i)\right\},
  $$ where $\nabla$ is the Laplacian on the group manifold $G$, and the $s_i$ are additional real variables.
  On top of representing, possibly, a unified framework for
  discrete quantum gravity approaches, this (class of) model(s) is
also closer to conventional field theories (even if still non
local) for the presence of derivatives in the action, and this
makes the analysis of its quantum (e.g. canonical) structure
easier, as we will discuss briefly in the following.

\section{The emergence of a continuum spacetime from Quantum Gravity: a different perspective and a research programme}

\

\subsection{The problem of the continuum from a GFT perspective}
Because the GFT formalism incorporates many ingredients that are
present in other approaches to quantum gravity, techniques that
have been developed to tackle the problem of the continuum within
them can be applied in GFT models as well. For example, one can
construct spin network weave states that approximate a given
semi-classical geometry at the kinematical level, as done in LQG,
insert them as boundary states (observables) in an appropriate GFT
model, use then the latter to define (using its tree level
perturbative expansion) the physical inner product among the
corresponding physical states, and finally compute observable
quantities that could be compared with the predictions of
continuum GR (or to experiments) to test the model. Alternatively,
one can use a GFT model with Feynman amplitudes given by the
exponential of a classical gravity action, truncate the sum over
geometric degrees of freedom in the perturbative expansion of the
GFT to reduce it to a purely combinatorial sum over
triangulations, possible further reduced to involve only trivial
topology (again, the tree level restriction is probably one way to
achieve this), and then apply the methods of the dynamical
triangulations approach: refinement and renormalization of the
free parameters of the GFT model. Finally, one can see the GFT
just as a definition of a particular spin foam model, to which one
can apply the methods for background independent coarse graining
developed in \cite{renorm}; this last option is particularly
revealing, because such methods have been originally developed as
a convenient and elegant encoding of perturbative renormalization
of ordinary QFTs, and indeed, from the perspective of GFTs, what
one is doing when applying those methods to spin foam models is
exactly performing perturbative renormalization moves to the
corresponding GFTs.

All this is very good, but it also suggests that there is much
more in the GFT formalism than what can be seen from the
perspective of the other approaches. In particular, the structures
that play such a prominent role in these approaches all appear at
the perturbative level in GFTs, and there is much more in a
quantum field theory that its perturbative expansion and a lot of
physics is either not captured at all or non very conveniently
described by it. Now let us instead take the GFT formalism
seriously. It seems to us that this means one thing: if quantum
spacetime is described (to some approximation and in some limited
context) by a specific class of GFTs, and if the emergence of a
continuum spacetime is a result of some non-perturbative physics
acting on a large number of its fundamental quantum discrete
constituents, then to work at the GFT perturbative level (in
whatever restriction corresponding to one of these other
approaches), is not the most convenient thing to do. We should
instead change perspective.

We propose to look at GFTs as fundamental quantum field theories
for the elementary quantum constituents of space, the fundamental
\lq atoms of space\rq mentioned in the introduction. These can be
seen, in GFTs, as simplices or as intertwiners, as discussed
above, and the theory describes their interaction in purely field
theoretic terms to form a discrete spacetime at the perturbative
level. Just as atoms of matter are described by quantum field
theories, these atoms of space would be described by group field
theories. Just as atoms of matter (or elementary particles), when
their number is limited, can be very well described by
perturbative QFT, in terms of Feynman diagrams and retain their
discrete appearance, the interactions of a few \lq atoms of
space\rq   is well-dealt with perturbatively in terms of spin foam
models, or of others discrete approaches, like simplicial quantum
gravity or causal sets, and gives rise to a discrete picture of a
quantum spacetime. At the same time, just as when the number of
atoms of matter is very large and their temperature is low, they
condense to a liquid (or some other condensed) phase and the field
theory picture in terms of elementary quanta gives way to an
effective hydrodynamic continuum description of the system, the
emergence of a continuum space(time) should be understood as the
condensation of a large number of interacting quantum atoms of
space to a new condensed phase, in which an hydrodynamic effective
description is more appropriate, but that can be deduced (at least
in part) from the underlying microscopic field theory. In other
words we are advocating a condensed matter picture for quantum
gravity, and we are suggesting that GFTs are the appropriate
context to realize this picture in detail, by taking them
seriously as microscopic field theories for the elementary quantum
constituents of space, i.e. the quantum gravity analog of the
field theory \ref{eq:He} for $^3 He$. We will now outline briefly
a possible research programme, whose goal is exactly to realize
the condensed matter picture we are advocating here and to solve
the problem of the emergence of a continuous classical spacetime
and of General Relativity as the effective description of its
geometry.

\subsection{The emergence of spacetime: a research programme}
Suppose now that we believe the proposed interpretation of Group
Field Theories, that we believe that the fundamental quanta of the
theory are indeed elementary atoms of space, and that the
continuum is to emerge from their interactions when their number
is large, after a phase transition from a \lq gas\rq phase to a
\lq liquid\rq phase, and that General Relativity is to appear as
the effective description of their collective dynamics in this
condensed phase. How should we proceed, starting from the GFT
action and partition function, once we understand this action and
partition function as the quantum gravity analog of \ref{eq:He},
to be dealt with using condensed matter physics tools and ideas?
The answer is obviously statistical field theory, i.e. we should
develop and study a statistical group field theory.

There are several techniques that are used in condensed matter
physics to deal with systems made of a large number of
constituents. One of them is the renormalization group, and in
fact one way to study the phase structure of GFTs would certainly
be in terms of exact renormalization group transformations. No
such study of GFT renormalization has been done until now, neither
at the non-perturbative level nor in perturbation expansion, and
it would certainly be of great value. This can be done using
Lagrangian setting, already available for GFTs. Here, however, we
would like to expand a bit more on an alternative strategy: the
development and use of a Hamiltonian statistical group field
theory. The development of such a picture requires two main steps.
1) {\it The Hamiltonian formulation of GFTs with the definition of
a clear Fock space structure} of the space of quantum states; this
has been developed partially in \cite{GFTfock} for the generalised
GFT models of \cite{generalised}, mentioned above, but that
analysis should be completed and then extended to other GFT
models. An interesting preliminary result of \cite{GFTfock} is
that the basic \lq atoms of space\rq are {\it fermions}, which is
important because, fermionic condensed matter systems show a
richer set of emergent phenomena in the low temperature regime. It
has to be checked whether this is confirmed for other GFT models.
2) One needs to {\it identify the GFT notion of temperature}.
Usually this is related to a Wick rotation (euclideanisation) of a
time variable. In the generalised models of \cite{generalised},
mentioned above, there is a multiplicity of time variables $s_i$,
a consequence of the non-local nature of GFTs, that is likely to
be present in any other model. A Wick rotation of these variables
is possible, but it is not obvious that this is what one should
look for, and in any case their multiplicity requires a
non-trivial adaptation of the traditional techniques. This time
multiplicity is also the main difficulty faced in the development
of an Hamiltonian formulation of GFTs. Also, if the notion of
temperature comes out of some of the variables of the GFT, as it
is expected, this would correspond from the quantum gravity point
of view to using some of the geometric degrees of freedom as
physical temperature. Once an Hamiltonian formulation and a notion
of temperature is obtained, the stage is set for the statistical
treatment of GFTs. The task will then be to analyze, with the
newly developed tools (RG methods and Hamiltonian methods), the
phase structure of various GFT models. The goal is to {\it prove
the existence of a liquid or generically condensed phase}, in
which only the collective behaviour of the atoms of space is
relevant. This will be by definition the phase of the theory where
a continuum approximation is appropriate, or in other words, where
a continuum spacetime emerges. The search for such a phase will be
done by studying the properties of the GFT system: i) {\it varying
the \lq number of atoms\rq}; ii) {\it varying the temperature of
the system} (one expect condensation in the low temperature
regime); iii) {\it varying type and strength of the interaction}
(which is also related, from the quantum gravity point of view, to
the strength of spatial topology changing processes; iv) {\it in
the GFT analogue of both the relativistic and non-relativistic
regimes} (from a purely formal point of view, the generalised GFT
models mentioned above have a non-relativistic Schoedinger-like
kinetic term in the time variables $s_i$); v) {\it varying of
course the type and symmetries of the fields and the group
manifold} on which they are defined (thus the specific GFT model
considered). Once the condensed/liquid phase has been found, and
thus the issue of the emergence of a continuum spacetime solved,
the the final goal is to {\it show the emergence of General
Relativity as the hydrodynamic description of the collective
variables in this phase}. Here one could build up directly on the
insights gained in condensed matter analogues of gravitational
phenomena \cite{volovik}.

The first thing to do is to {\it identify and study directly the
collective or hydrodynamic variables for the GFT system in the
condensed phase}. Given the microscopic GFT variables, one of them
is likely to be a connection field, which is also the right set of
variables for a gauge theory formulation of continuum gravity,
like those from which GFT have originated in the first place. The
argument for believing so is to look at what type of collective
variables emerge in the case of $^3 He$, for example: there one
has the steps: $\psi(v_i)\rightarrow \mid v_i^1,
v_i^2,...,v_i^N\rangle \rightarrow v_i(x)$, where the first step
is the Fock or multi-particle description of the system following
from the microscopic field theory with field $psi(v_i)$ one starts
from, and the last step is the passage to an hydrodynamic
continuous velocity field in the liquid phase. The analogues of
these steps in the GFT case would be something like:
$\phi(g_1,..,g_D) \rightarrow \phi(g_\mu)\rightarrow \mid g_\mu^1,
g_\mu^2,...,g_\mu^N\rangle \rightarrow g_\mu(x)\rightarrow
A^{IJ}_\mu(x)$, where $\mu = 1...D$ and the last step is the
correspondence between elementary (infinitesimal) parallel
transports and a connection field. Of course the exact nature and
number of the collective variables is model-dependent. Having done
so, the second step is to {\it derive the energy functional
(hamiltonian) or the Lagrangian} governing the effective dynamics
of these collective variables, and compare it with that of General
Relativity. The second way to proceed is based on {\bf the
analysis of quasi-particles}, i.e. the small excitations above the
ground state. One has first to {\it determine the
 energy spectrum} of these quasi-particles, which
characterizes the {\it universality class} of the condensed matter
system at hand, and can thus give immediate evidence of the
possibility of the emergence of gravity in the effective
description, One has then to {\it derive the equations of motion
and the effective Lagrangian for quasi-particle dynamics}, in the
vicinity of the Fermi points or of the Fermi surface, according to
the universality class to which the GFT belongs (if it is a
fermionic system); this will serve also to check that the
collective variables identified above are indeed seen by the
quasi-particle degrees of freedom as an effective metric field.
Finally, one can use for example Sakharov method of {\it induced
gravity} \cite{volovik} and derive the effective action for the
collective variables by integrating out the fluctuations of the
quasi-particles modes, and once more compare it with that of
General Relativity.

\section{Conclusions}
We have outlined above one possible strategy to solve the problem
of the continuum in Quantum Gravity using Group Field Theory
methods. Other strategies can be followed and different methods
from those proposed above can of course turn out to be more suited
to the task and more effective. Only future work can tell. The
only message that we think it is useful to put forward, even if
only as a suggestion, at the present stage, is the following: we
may have already found the right (class of) microscopic theory
(theories) of the fundamental constituents of a quantum spacetime.
We have in fact at our disposal a quantum field theory of
simplicial geometry, describing the interaction of elementary
building blocks of space, and with a discrete spacetime emerging
from this interaction; a large number of approaches converged to
this class of models, thus supporting and motivating them with a
variety of partial results and arguments; also, as we have
discussed, GFTs incorporate many of the key structures of other
approaches to discrete quantum gravity, and can thus also be seen
as a framework in which to realize and complete the insights of
these other approaches. The analysis of the perturbative structure
of these models has already proven to be fruitful and interesting,
but the probably much richer non-perturbative structure of them is
basically un-explored territory. In particular, we have argued in
the present contribution that GFTs may be the right framework in
which to realize the idea of spacetime as a condensate and of
General Relativity as an emergent effective theory for the
collective behaviour of the atoms of space in this phase, i.e. as
a geometro-hydrodynamics \cite{hu}, solving in this way the
outstanding problem of the continuum approximation that all
current discrete quantum gravity approaches still face. We do not
know at present if GFTs can really be all this; once more, only
much more future work can tell. For now, we only have to take them
seriously and start exploring.

\

\end{document}